\documentclass[showpacs,showkeys,aps,prl,twocolumn,superscriptaddress]{revtex4}
\usepackage{graphicx}
\usepackage{pstricks}
\usepackage{amsmath,amssymb}
\usepackage{epsfig}
\usepackage{dcolumn}

\newcommand{\ignore}[1]{}

\newcommand{\mc}[2]{\multicolumn{1}{#1}{#2}}
\newcommand{\mcc}[1]{\multicolumn{1}{c}{#1}}

\newcommand{\bfR}{ {\bf R}} 

\begin{document}

\title{Defect Formation Energies without the Band-Gap Problem: 
            Combining DFT and \textit{GW} for
            the Silicon Self-Interstitial}
\author{Patrick Rinke}
\affiliation{Materials Department, University of California, Santa Barbara, CA 93106, USA}
\affiliation{Fritz-Haber-Institut der Max-Planck-Gesellschaft,
             Faradayweg 4--6, 14195 Berlin, Germany}
\author{Anderson Janotti}
\affiliation{Materials Department, University of California, Santa Barbara, CA 93106, USA}
\author{Matthias Scheffler}
\affiliation{Materials Department, University of California, Santa Barbara, CA 93106, USA}
\affiliation{Fritz-Haber-Institut der Max-Planck-Gesellschaft,
             Faradayweg 4--6, 14195 Berlin, Germany}
\affiliation{Chemistry Department, University of California, Santa Barbara, CA 93106, USA}
\author{Chris G. Van de Walle}
\affiliation{Materials Department, University of California, Santa Barbara, CA 93106, USA}

\begin{abstract}
We present an improved method to calculate defect formation energies that 
overcomes the band-gap problem of Kohn-Sham density-functional theory (DFT) and reduces the
self-interaction error of the local-density approximation (LDA) to DFT.
We demonstrate for the silicon self-interstitial that
combining LDA with quasiparticle energy calculations
in the $G_0W_0$ approach increases the defect formation energy of the neutral charge state 
by $\sim$1.1~eV, which is in
good agreement with diffusion Monte Carlo calculations 
(E. R. Batista {\it et al.} Phys. Rev. B {\bf 74}, 121102(R) (2006), 
 W.-K. Leung {\it et al.} Phys. Rev. Lett. {\bf 83}, 2351 (1999)). 
Moreover, the $G_0W_0$-corrected charge
transition levels agree well with recent measurements.
\end{abstract}


\maketitle

Defects often noticeably influence the electrical and optical properties of a 
material by introducing defect states into 
the band gap. Reaching a microscopic understanding of the physical and chemical properties 
of defects in solids has long been a goal of first-principles electronic structure methods. 
Probably the most widespread theoretical method in this realm today is 
density functional theory (DFT) in the 
local-density (LDA) and generalized gradient approximation (GGA),
but certain intrinsic deficiencies limit their predictive power. 
Artificial self-interaction and the absence of the derivative discontinuity in the
exchange-correlation potential  \cite{Godby/Schlueter/Sham:1986} present the most notable deficiencies in this context. They give, amongst other things, rise to the band-gap problem -- the fact that the band gap in LDA and GGA 
underestimates the quasiparticle gap \cite{Godby/Schlueter/Sham:1986,Rinke/etal:2005}.
In this Letter we show that the  band-gap problem in LDA/GGA not only affects the reliable computation 
of defect levels, but in certain cases (e.g. filled defect states in the band gap) also that of formation energies. 
We present a formalism for calculating formation energies of defects in solids that
combines LDA with quasiparticle energy calculations in the $G_0W_0$ approximation
\cite{Hedin:1965} to reduce the self-interaction error and to overcome the band-gap problem.
In some cases a heuristic ``scissor operator" approach may approximately correct the problem. However, particularly when the experimental answer is unknown, a more accurate method is needed.

\begin{figure}[t] 
   \includegraphics[width=0.99\linewidth]{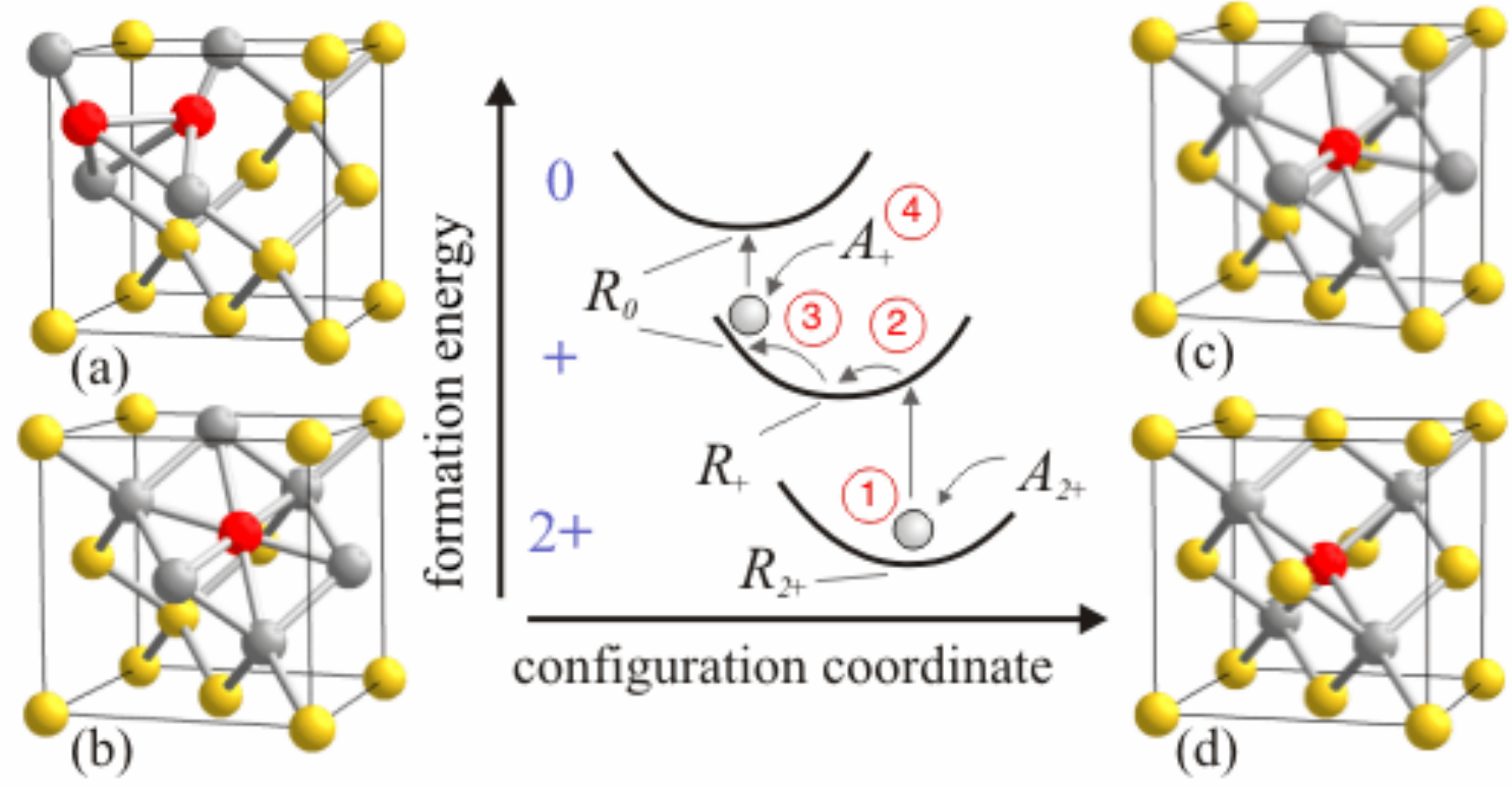}
   \caption{\label{fig:schem}(Color online)  a) Split$<$110$>$, b) hexagonal, c) $C_{3v}$ and d) tetrahedral
            configuration of the Si$_{\rm i}$. Defect atoms are shown in red and nearest neighbours in
	    grey. The middle panel depicts the formation of the neutral  Si$_{\rm i}$ from 
	    the 2+ charge state. $A_+$ and $A_{2+}$ are short for the electron affinities 
	    $A(+,\bfR_0)$ and $A(2+,\bfR_{2+})$ (see text), respectively, and $\bfR_q$ denotes the atomic positions
	    in charge state $q$.
	    }
\end{figure}

We illustrate our approach with the example of a self-intersitital in 
silicon (Si$_{\rm i}$), a defect of high technological relevance 
\cite{Bracht:2007,Coleman/Burrows:2007,Shimizu/Uematsu/Itoh:2007}.
In the neutral charge state the Si$_{\rm i}$ has several stable and metastable atomic 
configurations \cite{BarYam/Joannopoulos_1:1984,Car/etal:1984} (see
Fig.~\ref{fig:schem}), in all of which two electrons occupy a defect level in the band gap.
The LDA formation energies of all these configurations are underestimated severely (by  
$\sim$1.5 eV) compared to diffusion Monte Carlo (DMC) 
calculations \cite{Leung/Needs/etal:1999,Batista/etal:2006}. However, no insight into this
discrepancy is provided by the DMC calculations.

In our formalism the formation of the
neutral defect is expressed as successive charging of its 2+ charge state, 
for which the 
defect level is unoccupied. This allows us to decompose the formation energy into that of
the 2+ state ($E^f(2+)$), a lattice and an electron addition part.
This decomposition is not only insightful for analyzing the
underestimation of the LDA formation energy, but also permits us to 
apply the most suitable method for each type of contribution  \cite{Hedstroem/etal:2006}. 
For the lattice part we retain the LDA and argue
that the relaxation energies and $E^f(2+)$ are not as strongly affected by the deficiencies of the
LDA as in the positive and the neutral case since the defect level in the band gap is
unoccupied.
For the electron affinities, on the other hand, we employ Hedin's $GW$ method \cite{Hedin:1965}.
Since self-consistency in $GW$ is still discussed controversially 
\cite{Rinke/etal:2005} we obtain
the quasiparticle corrections to the LDA Kohn-Sham energies from 
first order perturbation theory ($G_0W_0$),
which is currently the
method of choice for computing quasiparticle band structures in solids
\cite{Aulbur/Jonsson/Wilkins:2000,Rinke/etal:2005}. 
While not completely self-interaction
free \cite{Nelson/Bokes/Rinke/Godby} $G_0W_0$ significantly reduces the self-interaction error. 
With this combined approach the formation energy in the neutral charge state 
increases by $\sim$1.1 eV compared to the 
LDA. Recent DFT  calculations employing the HSE hybrid functional, 
which also significantly reduces the self-interaction error, yield a similar improvement
\cite{Batista/etal:2006} and lend further substance to this notion.
Moreover, the $G_0W_0$-corrected charge
transition levels agree well with recent
experimental measurements \cite{Bracht:2007}.

\begin{figure}[t] 
   \includegraphics[width=0.99\linewidth]{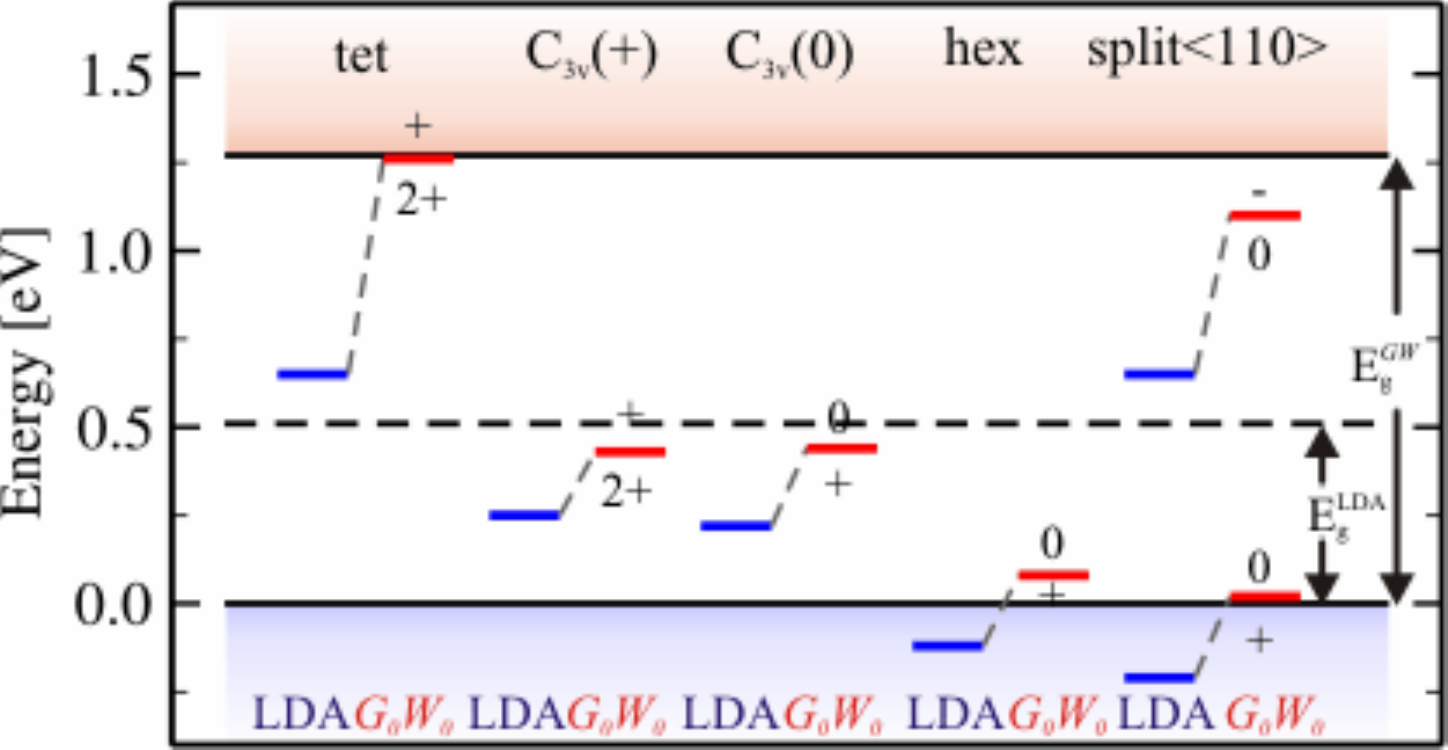}
   \caption{\label{fig:levels}(Color online)  
            Vertical electron affinities for different configurations of 
	    the Si$_{\rm i}$: LDA Slater transition states (blue) and $G_0W_0$
	    quasiparticle energies (red). 
	     }
\end{figure}

We will present our combined DFT+$G_0W_0$ approach for the example of the 
Si$_{\rm i}$, but it can easily be generalized to defects in other materials.
An additional silicon atom in an interstitial site can adopt different 
configurations with similar formation energies (cf. Fig. \ref{fig:schem}).
In the tetrahedral (tet) geometry the extra silicon atom gives rise to an $a_1$ and a 
threefold degenerate $t_2$ state. The latter is empty and in resonance with the 
conduction band. The partial occupation of the degenerate $t_2$ state triggers
a Jahn-Teller distortion along the $<$111$>$-axis into a geometry with $C_{3v}$ symmetry, also referred to as 
``displaced hexagonal structure'' in previous studies \cite{Al-Mushadani/Needs:2003}. 
The addition of a second electron 
displaces the atom further in the $<$111$>$-direction stabilizing the neutral 
charge state. This sequence is illustrated 
in Fig.~\ref{fig:levels}. Moving the interstitial atom along the $<$111$>$-direction
into the center of a six-membered ring (hexagonal (hex) configuration) lowers the neutral
state further in energy. It reaches its lowest position in the 
split$<$110$>$ configuration, where the added atom and a host atom share an
interstitial site oriented in the $<$110$>$-direction.

For the 2+ charge state the tetrahedral is the most stable configuration \cite{Car/etal:1984} (see also
Table~\ref{tab:data}) and we will use it as a starting point for building our  scheme. 
The positive charge state is then formed by adding one electron as depicted in steps 1 and 2
in Fig.~\ref{fig:schem}.
Mathematically this can be expressed by starting from the expression for the formation energy 
in the positive charge state
\begin{equation}
\label{eq:def_for_+}
   E_D^f(+,\epsilon_F)=E(+,\bfR_{+}^{D}) -E_{ref}+\epsilon_F \quad .
\end{equation} 
$E(q,\bfR_{q'}^D)$ is the total energy in charge state $q$ and atomic positions $\bfR_{q'}^D$ 
of defect configuration $D$ in charge state $q'$. $E_{ref}$ is a suitably chosen reference system, here bulk
silicon, and $\epsilon_F$ the Fermi level of the electrons referenced to the top of the valence
band. 
Adding and substracting first $E(+,\bfR_{2+}^{\rm tet})$ and then $E(2+,\bfR_{2+}^{\rm tet})$ leads to
\begin{alignat}{1}
\label{eq:def_for_+_++}
  E_D^f(+,\epsilon_F)=& \Delta(+,\bfR_+^D,\bfR_{2+}^{\rm tet})+A(2+,\bfR_{2+}^{\rm tet}) \nonumber \\
                      & +E_{\rm tet}^f(2+,\epsilon_F=0)
                      +\epsilon_F \quad .
\end{alignat}
The energy difference $E(+,\bfR_{2+}^{\rm tet})-E(2+,\bfR_{2+}^{\rm tet})$ defines the 
vertical electron affinity $A(2+,\bfR_{2+}^{\rm tet})$ of the 2+ state 
(in its tetrahedral configuration), step 1 in Fig.~\ref{fig:schem},
referenced to the top of the valence band, whereas 
$E(+,\bfR_+^D)-E(+,\bfR_{2+}^{\rm tet})$ gives the subsequent 
relaxation energy $\Delta(+,\bfR_+^D,\bfR_{2+}^{\rm tet})$ in the 
positive charge state (step 2). 

Similarly the neutral charge state emerges from the positive one by addition of an electron.
Mathematically we again achieve this by adding and substracting first 
$E(+,\bfR_0^D)$ and then $E(+,\bfR_+^D)$ to and from the
expression for the neutral formation energy
$E_D^f(0,\epsilon_F)=E(0,\bfR_0^D)-E_{ref}$: 
\begin{alignat}{2}
\label{eq:def_for_0_+}
  E_D^f(0,\epsilon_F)=& A(+,\bfR_0^D)+\Delta(+,\bfR_0^D,\bfR_+^D) \nonumber \\
                      &+E_f^D(+,\epsilon_F=0) \quad . 
\end{alignat}
Again $A(+,\bfR_0^D)$ denotes a vertical electron affinity 
$E(0,\bfR_0^D)-E(+,\bfR_0^D)$ (step 4) and 
$\Delta(+,\bfR_0^D,\bfR_+^D)=E(+,\bfR_0^D)-E(+,\bfR_+^D)$ the relaxation energy 
from the neutral to the positive geometry in the positive charge state (step 3).
An expression for the negative charge state can be obtained completely analogously 
once $E_D^f(0,\epsilon_F=0)$ has been computed.

The decomposition in Eq. (\ref{eq:def_for_+_++}) and (\ref{eq:def_for_0_+}) is not only appealing
from an intuitive point of view, but also groups the required total-energy 
differences into two categories: lattice
contributions in a fixed charge state and electron addition energies at fixed geometry.
This permits us to go beyond a pure DFT description in an easy fashion by
employing the most suitable method for each type of contribution \cite{Hedstroem/etal:2006}. 
Since we expect relaxation energies in the same charge state to be given reliably by LDA we
retain DFT for the lattice part. 
For the electron addition energies, i.e., changes in charge state, which are typically
problematic in LDA, we instead resort to the $G_0W_0$ method.

The last remaining quantity to be assigned is 
$E_{\rm tet}^f(2+,\epsilon_F=0)$, which we compute in the LDA. Unlike for the neutral state, 
the absence of DMC reference data unfortunately does not permit an assessment of the LDA error 
in this case. 
However, since the conduction-band-derived defect levels are unoccupied 
the effect of the self-interaction and the band-gap error on the formation energy 
should be small. 
We therefore expect LDA to be more reliable for the tetrahedral 2+ state than for the
neutral or the positive states. 

\begin{table*}
\begin{ruledtabular}
\begin{tabular}{l|d|dd||d|dd|dd}
 $D$ & \mc{c|}{$A(+,\bfR_0^D)$} 
     & \mcc{$\Delta(+,\bfR_+^D,\bfR_{2+}^{\rm tet})$} 
     & \mc{c||}{$\Delta(+,\bfR_0^D,\bfR_+^D)$} 
 & \multicolumn{1}{c|}{$E_D^f(2+)$}  
 & \multicolumn{2}{c|}{$E_D^f(+)$}
 & \multicolumn{2}{c}{$E_D^f(0)$} \\
 & & &   
        & \mc{c|}{LDA}  & \mcc{LDA} & \mc{c|}{$G_0W_0$} & \mcc{LDA} & \mc{c}{$G_0W_0$}   \\
	      \hline
  hex            & 0.08 & 0.402 & 0.012 &  3.73 & 3.41 & 4.31 & 3.40 & 4.40    \\
  split$<$110$>$ & 0.02 & 0.502 & 0.030 &  3.91 & 3.49 & 4.41 & 3.29 & 4.46    \\  
  C$_{3v}$       & 0.44 &-0.021 & 0.182 &  2.65 & 3.00 & 3.89 & 3.36 & 4.51  
\end{tabular}
\caption{\label{tab:data} 
         $G_0W_0$ vertical electron affinities  for 
	 different Si$_{\rm i}$ configurations $D$ and LDA relaxation energies. 
	 $\Delta(-,R_-^D,R_{0}^D)$ is -0.028~eV for the split$<$110$>$
	 and
	 $A(2+,\bfR_{2+}^{\rm tet})$ amounts to 1.26~eV in $G_0W_0$.
	 The tetrahedral configuration is taken as the 2+ state of the C$_{3v}$. 
	 Corrections for charged supercells (see text) have been added.
	 All values are given in eV.}
\end{ruledtabular}
\end{table*}

%
%
%

The LDA calculations in the present work have been performed with 
the plane-wave, pseudopotential code \texttt{S/PHI/nX} \cite{SFHIngX}.
64-atom supercells  
were used throughout, unless otherwise noted. 
To remove the contributions arising from the homogeneous compensation
charge density that is added to charged supercell calculations
we have performed calculations for supercells with 64, 
216 and 512 atoms. In these the interstitial atom was placed in the tetrahedral (2+) and the 
$C_{3v}$ (+) position of a perfect (undistored) Si lattice.
Fitting the formation energies up to cubic
order in the inverse cell length and extrapolating to infinite length we obtain
corrections to the 64-atom cell 
of 0.17~eV and 0.04~eV for the 2+ and + state,
respectively. Our extrapolated formation energy for the unrelaxed tetrahedral 2+ state of
3.19~eV agrees well the 3.31~eV obtained by Wright and Modine for a slightly larger lattice constant
\cite{Wright/Modine:2006}.
With this correction the formation energy of the relaxed tetrahedral 2+
configuration amounts to 2.65~eV. 
Applying a recently developed improved correction scheme \cite{newcorr}
yields a corrected value of 2.66~eV, in excellent agreement with our
extrapolated value.

For the $G_0W_0$ calculations \cite{convpar}
we have employed the $G_0W_0$ space-time code \texttt{gwst} 
\cite{GW_space-time_method:1998,GW_space-time_method_enh:2000,
      GW_space-time_method_surf:2007}. 
For computational convenience we calculate the electron affinity of positive charge 
states ($A(+,R_0^D)$) by their inverse process, the electron removal from the neutral state, 
since no spin polarization or partially filled defect states are encountered then.  
Separate $G_0W_0$ calculations for bulk silicon yield a band gap of 1.27~eV in good agreement with
previous pseudopotential $G_0W_0$ calculations \cite{Aulbur/Jonsson/Wilkins:2000}.

The computed vertical electron affinities are shown in Figure ~\ref{fig:levels}. 
For comparison the LDA affinities calculated as Slater transition states
\cite{Slater-STS:1974} at half occupation have been included.
The $G_0W_0$ corrections for  the +/0 state are similar for the three configurations and 
relatively small ($\sim$0.2~eV). For states that in the LDA are in resonance with the 
conduction band, however, the $G_0W_0$ corrections are much more pronounced.
Since these states have a contribution from delocalized conduction-band states 
the delocalization error of the LDA   
\cite{Mori-Sanchez/Cohen/Yan:2008} leads to a breakdown of Slater transition state 
theory. 
The resulting severe underestimation of the vertical affinities 
is akin to the band-gap problem. In LDA the band gap $E_{g}=I-A$, 
where $I$ is the ionization potential and $A$ the electron affinity, is underestimated 
regardless of whether $I$ and $A$ are calculated as total energy differences or by 
Kohn-Sham eigenvalues, because the exchange-correlation functional is a continuous 
function of the electron density and therefore does not exhibit a derivative
discontinuity. Many-body perturbation theoy in the $GW$ approach, on the other hand, 
does not suffer from this problem.

Having identified the relevant electron affinities we can now return to the formation energies
in Eqs. (\ref{eq:def_for_+_++}) and (\ref{eq:def_for_0_+}). 
Table ~\ref{tab:data} shows that
already upon adding the first electron to the 2+ state 
we observe a large correction ($\sim$0.9~eV) for the formation energy of the 
positive state. 
This error subsequently carries over to the 
neutral charge state, and adding the second electron incurs a further increase. 
The $G_0W_0$-corrected formation energies are now on average 1.1~eV larger 
than in the LDA.
Since the quasiparticle shift of the empty defect state in the split$<$110$>$ configuration is
smaller than the band-gap opening the state is moved into the band gap  
($A(0,\bfR_0^{\rm split})$=1.1~eV). 
As a result the negative charge state becomes stable in $G_0W_0$, which is not the 
case in LDA, and has a formation energy of 5.53~eV.

For the neutral charge state our $G_0W_0$ corrected formation energies compare well with 
recent DMC calculations that find an average increase of $\sim$1.5~eV (with a statitistical
error bar of $\pm$0.09~eV) and  
DFT HSE hybrid functional calculations that significantly reduce the 
self-interaction error and yield an average increase 
of $\sim$1.2~eV \cite{Batista/etal:2006}.
Earlier DMC calculations give a larger average increase of 1.7~eV compared to the LDA, but also
a much larger statistical error bar ($\pm$0.48~eV) \cite{Leung/Needs/etal:1999}.
Assuming a migration barrier of $\sim$0.2~eV \cite{Leung/Needs/etal:1999} our
computed activation enthalpy (formation energy + migration barrier) of $\sim$4.7~eV
for the neutral split$<$110$>$ interstitial is also in very good agreement with the
experimentally determined value of 4.95~eV \cite{Bracht:1995}.
 
\begin{figure}[t] 
   \includegraphics[width=0.99\linewidth, viewport=32 50 492 314]{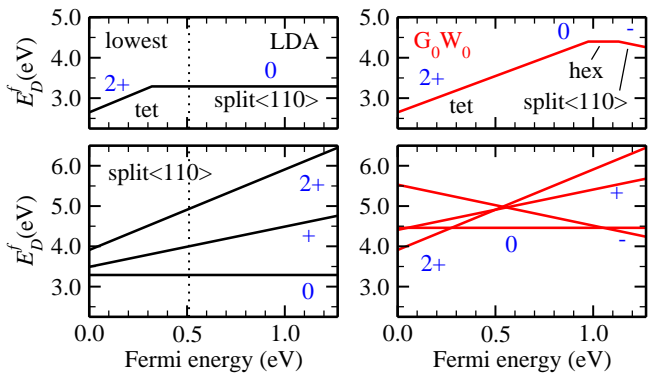}
   \caption{\label{fig:CTL} (Color online) Formation energies ($E_D^f$) as a function of 
            Fermi energy in LDA (left) and $G_0W_0$ (right). The lower panels show the
	    split$<$110$>$ as representative configuration and the upper
	    the configuration with the lowest energy for a give Fermi level. 
	    The dotted line marks the LDA band gap. 
	    }
\end{figure}

Finally we address the stability of the different defect configurations when the Fermi 
energy is varied  throughout the band gap (cf Fig.~\ref{fig:CTL}). For clarity this is 
shown only for the split$<$110$>$ configuration (lower panels) and the configuration with
lowest energy at a given Fermi level (upper panels). The situation for the 
hex and $C_{3v}$ configurations is qualitatively similar to that of the split$<$110$>$.
If each configuration is considered separately, the formation energy diagram looks 
startlingly different in LDA and $G_0W_0$. Since LDA underestimates the formation
energies of the + and 0 state relative to the 2+ it does not exhibit the
negative-$U$ behavior 
($E_D^f(+) > \{E_D^f(0),E_D^f(2+)\}$ for all Fermi energies)  
observed in $G_0W_0$.
In addition $G_0W_0$ stabilizes the negative charge state for the 
split$<$110$>$ interstitial.
If, on the other hand, the configuration with the lowest energy is
considered, LDA and $G_0W_0$ superficially give a more similar picture: 
the tetrahedral 2+ state is stable for 60-70\%
of the respective band gaps. While LDA then gives preference to the neutral split$<$110$>$ for
larger Fermi levels,  the $G_0W_0$ corrections marginally stabilize the neutral
hex configuration, in agreement with the earlier DMC calculations
\cite{Leung/Needs/etal:1999}. The actual energies and transition levels between LDA and
$G_0W_0$, however, differ appreciably.


Every point at which two lines in Fig.~\ref{fig:CTL} cross corresponds to a charge-state 
transition level $\varepsilon_{q/q'}$. 
Bracht {\it et al.} have recently determined these for the
silicon self-interstitial in high temperature diffusion experiments \cite{Bracht:2007}. They 
identified two levels, at $\approx$ 0.1-0.2~eV and at $\approx$ 0.4~eV 
above the valence-band maximum, that they ascribed to $\varepsilon_{0/+}$ and 
$\varepsilon_{+/2+}$, respectively. 
These would most
closely correspond to the $G_0W_0$-corrected  charge-state transition levels  
$\varepsilon_{0/+}$=0.09~eV and $\varepsilon_{+/2+}$=0.58~eV for the hexagonal 
configuration or
$\varepsilon_{0/+}$=0.05~eV and $\varepsilon_{+/2+}$=0.50~eV for the split$<$110$>$, 
while those of the $C_{3v}$ configuration are noticeably different
(0.62~eV and 1.24~eV). 
Although lowest in formation energy and therefore 
highest in concentration, the $C_{3v}$ 2+ configuration (which is identical to tet
2+) most likely plays a negligible 
role in the diffusion experiments, since
its diffusion would have to proceed through a hexagonal site. The activation energy for this
process (formation energy + energy barrier at the experimental situation of 
a Fermi level close to the middle of the band gap \cite{Bracht:2007}) 
would thus be considerably larger than the activation energy for diffusion processes involving the other configurations.
Refinements in the diffusion models (e.g. inclusion of multiple configurations and charge-state dependent diffusion barriers) may be able to clarify the role of the tet 2+ configuration in future experimental studies.


We gratefully acknowledge fruitful discussions with K. Delaney, C. Freysoldt, J. Neugebauer 
and E. K. U. Gross.
This work was supported by the NSF MRSEC Program under award No. DMR05-20415 and 
the Nanoquanta Network of Excellence (NMP4-CT-2004-500198). 
P. Rinke acknowledges the Deutsche Forschungs Gemeinschaft, 
the UCSB-MPG Exchange and the NSF-IMI Program (DMR04-09848) for financial support.
Some computations were performed at the San Diego Supercomputer
Center under grant number DMR070072N.


\begin{thebibliography}{24}
\bibitem[{con()}]{convpar}
\bibinfo{note}{A plane-wave cutoff of 32~Ry (14~Ry) and a $3\times3\times3$
  ($2\times2\times2$) $\Gamma$-point centered Monkhorst-Pack (MP) k-point mesh
  have been used for the exchange (correlation) part of the $G_0W_0$
  self-energy. Head and wings of the dielectric matrix have been preconverged
  on a $4\times4\times4$ MP off-center mesh}.
\end{thebibliography}

\end{document}